\documentclass[twocolumn,preprintnumbers,nofootinbib,prd,superscriptaddress,aps]{revtex4-1}

\usepackage[utf8]{inputenc} 
\usepackage{graphicx,amssymb,amsmath,amsthm,amsfonts,epstopdf,epsfig,epsf,times}
\usepackage[linktocpage]{hyperref}
\usepackage[usenames]{color}
\usepackage{epstopdf}
\usepackage{textcomp}
\usepackage{bm}
\usepackage{latexsym}
\usepackage{rotating}
\usepackage{hyperref}
\usepackage{color}
\usepackage{longtable}
\usepackage{enumerate}
\usepackage{tensor}
\usepackage{stmaryrd}
\usepackage[normalem]{ulem}
\usepackage{mathtools}
\usepackage{url}
\usepackage{multirow}
\usepackage{graphicx}

\definecolor{coolblack}{rgb}{0.0, 0.18, 0.39}
\definecolor{darkred}{rgb}{0.5,0,0}
\definecolor{darkgreen}{rgb}{0,0.5,0}
\definecolor{darkblue}{rgb}{0,0,0.5}
\definecolor{lapislazuli}{rgb}{0.15, 0.38, 0.61}
\definecolor{venetianred}{rgb}{0.78, 0.03, 0.08}
\definecolor{bleudefrance}{rgb}{0.19, 0.55, 0.91}
\definecolor{dogwoodrose}{rgb}{0.84, 0.09, 0.41}
\definecolor{dogwoodrose}{rgb}{0.84, 0.09, 0.41}
\definecolor{darkorgane}{rgb}{1,0.549,0}
\hypersetup{colorlinks=true, citecolor=darkblue, linkcolor=darkblue, 
urlcolor = darkblue}
\definecolor{olive}{rgb}{0.5, 0.5, 0.0}

\newcommand{\ben}{\begin{enumerate}}
\newcommand{\een}{\end{enumerate}}

\newcommand{\del}{\partial}

\def\be{\begin{equation}}
\def\ee{\end{equation}}
\newcommand{\beq}{\begin{eqnarray}}
\newcommand{\eeq}{\end{eqnarray}} 
\newcommand{\ba}{\begin{align}}
\newcommand{\ea}{\end{align}}

\def\be{\begin{equation}}
\def\ee{\end{equation}}
	
\newcommand{\bea}{\begin{eqnarray}}
\newcommand{\eea}{\end{eqnarray}}

\begin{document}
\title{Environmental effects in gravitational-wave physics: tidal deformability of black holes immersed in matter}
\author{Vitor Cardoso}
\affiliation{CENTRA, Departamento de F\'{\i}sica, Instituto Superior T\'ecnico -- IST, Universidade de Lisboa -- UL,
Avenida Rovisco Pais 1, 1049 Lisboa, Portugal}
\affiliation{Institute for Theoretical Physics, University of Amsterdam, PO Box 94485, 1090GL, Amsterdam, The Netherlands}
\author{Francisco Duque}
\affiliation{CENTRA, Departamento de F\'{\i}sica, Instituto Superior T\'ecnico -- IST, Universidade de Lisboa -- UL,
Avenida Rovisco Pais 1, 1049 Lisboa, Portugal}

\begin{abstract} 
The tidal deformability of compact objects by a companion has a detectable imprint in the gravitational waves emitted by a binary system.
This effect is governed by the so-called tidal Love numbers. For a particular theory of gravity, these depend solely on the object internal structure and they vanish for black holes in general relativity. A measurement compatible with nonzero tidal Love numbers could thus provide evidence of new physics in the strong-field regime. However, in realistic astrophysical scenarios, compact objects are surrounded by a nonvacuum environment. In this work, we study the tidal deformability of configurations of black holes immersed in matter, focusing on two analytical models representing an anisotropic fluid and a thin-shell of matter around a black hole. We then apply our results to the astrophysically relevant case of a black hole surrounded by an accretion disk, in the parameter region of interest of the upcoming LISA mission. Our results indicate that there are challenges to overcome concerning tests of strong-field gravity using tidal Love numbers.
\end{abstract}
\maketitle
\section{Introduction} \label{sec:Intro}

Tidal interactions are responsible for many astrophysical phenomena that have caught our attention since the dawn of Newton's theory of gravitation~\cite{poisson_will_2014}. The most obvious example are ocean tides, caused by differences in the gravitational field produced by the Moon or Sun at different places on the Earth. Their tidal interaction also explains why the Earth is losing angular momentum/rotational energy to the Moon and why the days are getting longer. More recently, violent events have been observed at an astrophysical scale caused by tidal effects, such as tidal disruptions in close binary systems.

The tidal distortion of a compact object due to an external gravitational field is quantified, at linear level, using tidal Love numbers (TLNs)~\cite{poisson_will_2014}. They are the analogous of a gravitational susceptibility. The TLNs only depend on the dynamics of the gravitational field, i.e. the underlying theory of gravity, and the internal structure of the deformed body and appear in the orbital equation of motions of a binary system at leading Newtonian order~\cite{Mora:2003wt, Vines:2010ca}. They introduce corrections in the gravitational waveform emitted by a coalescing binary during the late-inspiralling phase, at 5 post-Newtonian order~\cite{Flanagan:2007ix, Hinderer:2016eia}. These prospects motivated a development of a relativistic theory of TLNs \cite{Binnington:2009bb, Damour:2009vw, Hinderer:2007mb}, in order to use gravitational-wave (GW) measurements to understand the structure of more compact objects. 

The first studies on the subject focused on the TLNs of neutron stars, and provide access to the equation of state above the currently understood nuclear densities~\cite{Baiotti:2019sew}. More recently, tidal deformations have been proposed to be a good candidate to test strong-field gravity, the black hole (BH) paradigm and to search for new exotic, compact objects~\cite{Cardoso:2017cfl,Maselli:2017cmm,Maselli:2018fay,Pani:2019cyc}. A crucial aspect in this research program is the fact that the TLNs of BHs vanish in General Relativity (GR)~\cite{Binnington:2009bb, Damour:2009vw, Gurlebeck:2015xpa,Pani:2015hfa,Landry:2015zfa} (at least to second order in spin). Thus, a measurement of a nonvanishing TLN is evidence for new physics: either the object is not a BH, or GR is not the most accurate description of gravity.

Consider the first possibility above. Quantum corrections at the horizon scale, or exotic matter (or exotic equations of state) could lead to the formation of exotic compact objects (ECOs)~\cite{Cardoso:2019rvt}. 
When the BH limit is approached,
\beq
\mathcal{C}\coloneqq M/r_0\rightarrow 1/2 \, ,
\eeq
where $M$ and $r_0$ are, respectively, the mass and radius of the ECO, its TLNs generically converge to the BH limit (zero), but 
(for many models) logarithmically~\cite{Cardoso:2019rvt}.
Finally, using a Fischer matrix approach to estimate the errors on the TLNs, one concludes that ground-based detectors like Advanced LIGO and the Einstein Telescope can only set constraints on low compact ECOs, while upcoming space-interferometer LISA would probe the tidal deformability of ECOs almost up to the BH limit~\cite{Maselli:2017cmm,Maselli:2018fay,Pani:2019cyc,Cardoso:2019rvt}.
Nonzero TLNs may signal corrections to GR. Extra degrees of freedom create extra tidal fields for which a theory of TLNs is still poorly formulated~\cite{Bernard:2019yfz}. For some of the extensions considered, and setting to zero the asymptotic tidal moments of the extra fields, BHs do have nonzero TLNs~\cite{Cardoso:2018ptl,Cardoso:2019rvt}.

There is, however, a third unexplored option which could be responsible for the (apparent) non vanishing of TLNs of a BH: the presence of external matter. Previous works on tidal deformability assumed pure vacuum environments (apart from the matter constituting the compact object). However, any astrophysically plausible self-gravitating object will be surrounded by matter. Such composite system will
in general have small but nonzero effective TLNs, limiting our ability to understand the nature of the compact object or of the underlying theory of gravity.
Thus, in order to use tidal deformability to test GR and the nature of compact objects, it is imperative to quantify the effects of this \textit{environment}. 

This is precisely the aim of our work. The picture we have in mind is that of a coalescing binary immersed in matter. The TLNs of each isolated BH are zero but the surrounding matter will also be deformed by any external field and, in principle, the TLNs of the configuration BH \textit{and} surrounding matter will not vanish.  

This paper is organized as follows. In Sec.~\ref{sec:FormalismTLN}, we review the general theory of tidally deformed self-gravitating objects in GR, in spherically symmetric, static background spacetimes. Using this formalism, we compute in Sec.~\ref{sec:DirtyBH} the TLNs of two spacetimes modeling a BH immersed in matter. In Sec.~\ref{sec:EoM}, we discuss the portion of environmental matter which is relevant to the dynamics of the coalescing binary. This analysis provides a typical length scale for the matter surrounding a BH which plays a role in tidal deformability. Finally, in Sec.~\ref{sec:Astro} we model the environment in which the binary is immersed as a Shakura-Sunayev thin disk~\cite{Shakura:1972te} and apply the results obtained in previous sections to astrophysical scenarios of potential interest to LISA.

\section{Setup: tidal deformability in General Relativity} \label{sec:FormalismTLN}
We start by reviewing the general theory of tidally deformed objects in GR~\cite{Hinderer:2007mb,Binnington:2009bb,Damour:2009vw,Cardoso:2017cfl}. Consider an isolated, self-gravitating compact object to which we add an external tidal field. The latter can be described in terms of tidal multipole moments of order $l$~\cite{Binnington:2009bb,Cardoso:2017cfl}
\beq
	\mathcal{E}_{a_1...a_l}&\equiv& \left[\left(l-2\right)!\right]^{-1}\left<C_{0a_1;a_3...a_l}\right> \, , \\
	\mathcal{B}_{a_1...a_l}&\equiv& \left[\frac{2}{3}\left(l+1\right)\left(l-2\right)!\right]^{-1}\left<\epsilon_{a_1bc}C^{bc}_{a_20;a_3...a_l}\right> \, ,
\eeq
where $C_{abcd}$ is the Weyl tensor, the semicolon denotes covariant derivative, $\epsilon_{abc}$ is the permutation tensor and the angular brackets denote symmetrization and trace removal. $\mathcal{E}_{a_1...a_l}$ ($\mathcal{B}_{a_1...a_l}$) are the polar (axial) moments and can be expanded in a basis even (odd) parity spherical harmonics $Y^{lm}\left(\theta,\varphi\right)$ defined by
\beq
Y^{lm}(\theta,\varphi)\equiv\sqrt{\frac{2l+1}{4\pi}\frac{\left(l-m\right)!}{\left(l+m\right)!}}P^{m}_l\left(\cos\theta\right)e^{im\varphi}\, .
\label{eq:SphericalHarmonic}
\eeq

The tidal field perturbs the equilibrium configuration of the previously isolated compact object. This deformability can be described by a change in its multipole moments. If we treat the tidal field as a small perturbation, we can use standard linearized GR to compute this change. We start by perturbing the background spacetime which describes the compact object
\beq
g_{\mu\nu}=g_{\mu\nu}^{\left(0\right)} + h_{\mu\nu} \, , \label{eq:FullMetric}
\eeq
where $g_{\mu\nu}^{\left(0\right)}$ is the background spacetime metric and $h_{\mu\nu}$ is a small perturbation. Throughout this work, we focus exclusively on spherically symmetric, static backgrounds whose general form is given by 
\be
ds^2= -F\left(r\right)dt^2 + G\left(r\right) dr^2 + r^2 d\theta^2 + r^2 \sin^2\theta d\varphi^2 \label{eq:sphericalmetric}\, .
\ee

In order to exploit the spherical symmetry of the background, we expand $h_{\mu\nu}$ in spherical harmonics and separate them in even and odd components. In the Regge-Wheeler gauge \cite{Regge:1957td}, they can be written as
\begin{widetext}
\beq
h_{\mu\nu}^{\text{even}}&=&
\begin{pmatrix}
F\left(r\right)\, H_0^{lm}\left(t,r\right)\,Y^{lm}& H_1^{lm}\left(t,r\right) \,Y^{lm} 		 & 0 & 0 \\
H_1^{lm}\left(t,r\right) \,Y^{lm} 				& 	G\left(r\right)\,H_2^{lm}\left(t,r\right) \,Y^{lm}	 & 0 & 0 \\
0 & 0 									&	r^2 \,K^{lm}\left(t,r\right) \,Y^{lm}	 & 0 \\
0 & 0 									&	0							 & r^2\sin^2\theta\, K^{lm}\left(t,r\right) \,Y^{lm}	
\end{pmatrix}\, ,\\
h_{\mu\nu}^{\text{odd}}&=&
\begin{pmatrix}
0 & 0		 & h_0^{lm}\left(t,r\right)\,S_\theta^{lm}  & h_0^{lm}\left(t,r\right)\,S_\varphi^{lm} \\
0 & 0 		 & h_1^{lm}\left(t,r\right)\,S_\theta^{lm}  & h_1^{lm}\left(t,r\right)\,S_\varphi^{lm} \\
h_0^{lm}\left(t,r\right)\,S_\theta^{lm} & h_1^{lm}\left(t,r\right)\,S_\theta^{lm} & 0 & 0\\
h_0^{lm}\left(t,r\right)\,S_\varphi^{lm}   & h_1^{lm}\left(t,r\right)\,S_\varphi^{lm}	& 0 & 0
\end{pmatrix} \, ,
\label{eq:MetricPerturbations}
\eeq
\end{widetext}
where
\beq
\left(S_\theta^{lm},S_\varphi^{lm}\right)\equiv\left(-Y^{lm}_{,\varphi}/\sin\theta,\sin\theta \, Y^{lm}_{,\theta}\right) \,,
\eeq
and $Y^{lm}=Y^{lm} (\theta,\varphi)$ are scalar spherical harmonics.
Then, we solve the field equations to linear order in $h_{\mu\nu}$ for a specific background. When the perturbed object is nonrotating, the even parity sector decouples completely from the odd parity sector.

Now, we have to extract the tidal fields and induced multipole moments from the asymptotic behavior of the full metric \eqref{eq:FullMetric}. This can be done using Thorne's definition of multipole coefficients for any stationary, asymptotically flat spacetime using asymptotically Cartesian and mass centered coordinates~\cite{Thorne:1980ru}. An equivalent, coordinate independent definition of multipole moments for axisymmetric and asymptotically flat spacetimes was put forward by Geroch and Hansen \cite{Geroch:1970cd, Hansen:1974zz} (for a review see Ref.~\cite{Cardoso:2016ryw}). We use the following asymptotic expansions of the metric components
\begin{widetext}
\beq
	g_{tt}&=&-1+\frac{2M}{r}+\sum_{l\geq 2}\left( \, \frac{2}{r^{l+1}}\left[\sqrt{\frac{4\pi}{2l+1}}M_l\,Y^{l0}+\left(l'<l \, \text{pole}\right)\right]-\frac{2}{l\left(l-1\right)}r^l\left[\mathcal{E}_lY^{l0} +\left(l'<l \, \text{pole}\right) \right]   \right) \,,\\ \label{eq:PolarMetricExpansion}
	g_{t\varphi}&=&\frac{2J}{r}\sin^2\theta+\sum_{l\geq 2}\left( \, \frac{2}{r^{l}}\left[\sqrt{\frac{4\pi}{2l+1}}\frac{S_l}{l}\,S^{l0}_\varphi+\left(l'<l \, \text{pole}\right)\right]+\frac{2r^{l+1}}{3l\left(l-1\right)}\left[\mathcal{B}_lS^{l0}_\varphi +\left(l'<l \, \text{pole}\right) \right]   \right) \, ,\label{eq:AxialMetricExpansion}
\eeq
\end{widetext}
where $M_l$ are the mass multipole moments, $S_l$ are the current multipole moments, and $\mathcal{E}_l$ and $\mathcal{B}_l$ are, respectively, the amplitudes of the polar and axial components of the external field with harmonic number $l$, where axisymmetry was used to fix $m=0$.

Finally, we define the polar and axial TLNs, respectively, as the dimensionless ratios \cite{Cardoso:2017cfl}
\beq
	k_l^E&=&-\frac{1}{2}\frac{l\left(l-1\right)}{M^{2l+1}}\sqrt{\frac{4\pi}{2l+1}}\frac{M_l}{\mathcal{E}_{l0}} \, , \label{eq:PolarTLNs} \\
	k_l^B&=&-\frac{3}{2}\frac{l\left(l-1\right)}{\left(l+1\right)M^{2l+1}}\sqrt{\frac{4\pi}{2l+1}}\frac{S_l}{\mathcal{B}_{l0}}  \,  , \label{eq:AxialTLNs}
\eeq
where $M$ is the mass of the object. Most references~\cite{Damour:2009vw, Binnington:2009bb, Hinderer:2007mb} normalize the TLNs in powers of the object radius $R$ instead of $M$, because they were working with bodies with a hard surface, e.g. neutron stars. Here, we adopt the convention of Ref.~\cite{Cardoso:2017cfl} since the radius of distributions of matter surrounding BHs is generally ill defined, as occurs for some ECOs. Consequently, the two definitions are related by
\beq
k_{l\,\text{ours}}=\left(\frac{R}{M}\right)^{2l+1}k_{l\,\text{standard}}\,.
\label{eq:TLNdiff}
\eeq

We should mention that the invariant character of this definition of TLNs has been put into question.~\cite{Gralla:2017djj}. Nonetheless, we are interested in comparing TLNs of different distributions of matter within a fixed coordinate system, and we will never interpret the TLNs obtained in terms of their singular value. As shown in the same work~\cite{Gralla:2017djj}, there is no ambiguity in this procedure and thus on the conclusions obtained on the following sections.

\section{Black holes surrounded by matter\label{sec:DirtyBH}}

\subsection{Black holes with short hair}\label{sec:ShortHair}

We consider two models for BHs surrounded by matter fields. The first model is that of a static, spherically symmetric spacetime containing an anisotropic fluid~\cite{Brown:1997jv}. The following line element -- of the form \eqref{eq:sphericalmetric} -- describes a BH surrounded by a fluid which satisfies both the weak and strong energy condition,
\beq
F\left(r\right)&=&G^{-1}\left(r\right)= 1 - \frac{2M}{r} - \frac{Q^{2k}}{r^{2k}} \,,\\
\rho &=& \frac{Q^{2k}\left( 2k-1 \right)}{8 \pi r^{2k+2}} \quad , \quad P = k\rho \label{eq:shorthairmatter}\, ,
\eeq
where $\rho$ and $P$ are, respectively, the matter density and the pressure on the isotropic $\theta$-$\phi$ surfaces, while $Q$ is a constant. The energy-momentum tensor of the fluid is
\be
T_{\mu\nu}=\rho \left( U_\mu U_\nu - u_\mu u_\nu \right) + P \sigma_{\mu\nu} \, , \label{eq:StressTensorFluid}
\ee
where $U^\mu$ is the fluid's 4 velocity, while $u_\nu$ and $\sigma_{\mu\nu}$ are, respectively, the unit normal and metric of the isotropic 2-spheres
\beq
	U^\mu&=&\left(-\frac{1}{\sqrt{F\left(r\right)}},0,0,0\right) \, , \\
	u_\nu&=&\left(0,\sqrt{G\left(r\right)},0,0\right) \, , \\
	\sigma_{\mu\nu}&=&\text{diag}\left(0,0,r^2,r^2\sin^2\theta\right)\, .
\eeq
For $k=1$, this class of BHs yields the Reissner-Nordstr\"{o}m solution~\cite{Brown:1997jv}. For $k>1$, the parameter $Q$ corresponds to a \textit{matter-hair}~\cite{Brown:1997jv,Barausse:2014tra}, which can be arbitrarily short by taking $k$ to be arbitrarily large.

To determine the TLNs of this configuration, we need to complement the gravitational perturbations \eqref{eq:MetricPerturbations} with the ones from the matter sector, where any equilibrium background quantity $X=X_0$ gets perturbed by the external tide, $X\to X_0+\delta X(t,r,\theta,\varphi)$.
%
%
We consider the regime of static tides, where the time variations of the tidal field are small compared to the dynamical timescale of the system. Consequently, all the perturbations introduced are independent of the coordinate time $t$. This immediately fixes $U^\mu$ and $u_\mu$ by imposing the correct normalizations ($U^2=-1$ and $u^2=1$), and that $U^\mu$ remains proportional to the timelike killing vector field $\del/\del t$
\beq
	\delta U^\mu&=&\left(\frac{1}{2\sqrt{F}}H_0^{lm}Y^{lm},0,0,0\right) \, , \\
	\delta u_\mu&=&\left(0,\frac{\sqrt{G}}{2}H_2^{lm}Y^{lm},0,0\right) \, .
\eeq
For $\sigma_{\mu\nu}$, we allow one more degree of freedom that respects the background spherical symmetry
\be
\delta \sigma_{\mu\nu}=\text{diag}\left(0,0,r^2 \, K_2^{lm}(r) \, Y^{lm},r^2\sin^2\theta\, K_2^{lm}(r)\,Y^{lm}\right)	\,.
\ee	
Finally, we let 
\beq
\rho&=&\rho_0+\delta \rho^{lm}\left(r\right)\,Y^{l\,m} \, , \\
P&=&P_0+\delta P^{lm}\left(r\right)\,Y^{lm}\,.
\eeq
%
\subsubsection{Axial perturbations}
The axial sector of stationary gravitational perturbations is entirely decoupled from matter perturbations~\cite{Damour:2009vw,Binnington:2009bb, 1968ApJ...152..673T}. Consequently, the $t\varphi$-component of Einstein's equations yields a decoupled ordinary differential equation for $h_0$,
\beq
&&r^2\left(1-\frac{2M}{r} + \frac{Q^{2k}}{r^{2k}}\right)h_0'' \nonumber \\
&=&\left(l\left(l+1\right)-\frac{4M}{r}+2k \left(1+2k\right) \frac{Q^{2k}}{r^{2k}}\right)h_0  \, .
\label{eq:Aniaxialmaster}
\eeq

We now follow a similar approach to that in Ref.~\cite{Cardoso:2017cfl}, by treating the matter-hair perturbatively. As such, we expand the metric perturbations in powers of the adimensionalized coupling $\epsilon=Q^{2k}/M^{2k}$,
\be
h_{\mu\nu}=h_{\mu\nu}^{\left( 0 \right)}+ \epsilon\, h_{\mu\nu}^{\left( 2 \right)} \, , \label{eq:expansionmatterhair} \, 
\ee
where $h_{\mu\nu}^{\left( 0 \right)}$ is the vacuum GR solution. For $l=2$, the $0th$-order axial perturbation regular at the horizon is
\be
h_0^{\left( 0 \right)}=\frac{\mathcal{B}_2}{3}r^3\left(1-\frac{2M}{r}\right)\, .
\ee
Expanding Eq.~\eqref{eq:Aniaxialmaster} to order $\mathcal{O}\left(\epsilon\right)$ we find
%
\beq
&&\left(\frac{d^2}{dr^2}+2\frac{2M-3r}{r^2\left(r-2M\right)}\right)h_0^{\left( 2 \right)} =  \frac{2\mathcal{B}_2}{3}\left(\frac{M}{r} \right)^{2k}\nonumber\\
&\times&\frac{r\left(\left(2k^2+k-3\right)r-\left(2k^2+k-1\right)2M\right)}{r-2M}\, .\label{eq:Axialperturbation}
\eeq
This equation admits a solution in closed form, in terms of the homogeneous solution and an hypergeometric function.
From it, we read the TLNs,
%
%
%
%
%
%
%
\be
k_2^B=\frac{1}{5}\frac{2^{5-2k}\left(2k-1\right)}{2k^2-9k+10}\frac{Q^{2k}}{M^{2k}}\,,\quad k>2\,.\label{eq:AxialTLNs_fluid}
\ee
For $k=1$ we find $k_2^B=0$ as expected since one then recovers the charged BH solution~\cite{Cardoso:2017cfl}. For $k=2$, we find
a dominant logarithmic term $\log(r)/r^2$ which is new and for which we lack a physical interpretation.
We can of course express the above in terms of the mass $\delta M\sim Q^{2k}/M^{2k}M$ contained in the fluid: $k_2^B \sim \delta M/M$.

%
%
%
\subsubsection{Polar perturbations}
In the polar sector, matter perturbations are no longer decoupled from the gravitational ones and one needs to consider also an equation of continuity coming from the conservation of the energy momentum tensor~\eqref{eq:StressTensorFluid}. The $tr$-component and $\theta$$\theta$-component of Einstein's equations, respectively yield
\be
H_1=0 \,,\quad H_2=H_0\, .
\ee
The $\theta$-component of the Bianchi identity ($\nabla_a T^{ab}=0$) fixes the pressure perturbation to be
\be
\delta P=2k\left(2k-1\right)\frac{Q^{2k}}{r^{2k}} \frac{K-K_2}{16\pi r^2}\, .
\ee

The $tt$, $rr$, and the $\theta\theta$ component of Einstein's equations provide expressions for $K_0''$,$ K_0'$ and $K_0$ in terms of $H_0''$, $H_0'$, $H_0$. Substituting these in the $tr$-component of Einstein's equations gives the following decoupled ordinary differential equation for $H_0$
\begin{widetext}
\beq
&&r^2\left(1-\frac{2M}{r}+\frac{Q^{2k}}{r^{2k}} \right)^2 H_0''+2r\left(1-\frac{2M}{r}+\frac{Q^{2k}}{r^{2k}} \right)\left(1-\frac{M}{r}+\left(1-k\right)\frac{Q^{2k}}{r^{2k}} \right) H_0' \nonumber \\
&=&\left(l\left(l+1\right)+\frac{4M^2}{r^2}-2l\left(l+1\right)\frac{M}{r}+ \left(l\left(l+1\right)+2k\left(1-\frac{6M}{r}\right)-4k^2\left(1-\frac{2M}{r}\right) \right)\frac{Q^{2k}}{r^{2k}}+2k \frac{Q^{4k}}{r^{4k}}\right)H_0\, .
\label{eq:PolarFluid}
\eeq
\end{widetext}

Following the same approach as in the axial case, we treat the matter-hair as a perturbation to GR using the expansion in eq. \eqref{eq:expansionmatterhair}. For $l=2$, the polar perturbation regular at $r=2M$ is
\be
H_0^{\left(0\right)}=-\mathcal{E}_2\,r^2\left(1-\frac{2M}{r}\right)\, , 
\ee
and Eq.~\eqref{eq:PolarFluid} can be written as
%
\be
\left(\frac{d^2}{dr^2}+\frac{\left(r-M\right)}{r\left(r-2M\right)}\frac{d}{dr}-\frac{2\left(2M^2-6Mr+3r^2\right)}{r^2\left(r-2M\right)}\right)H_0^{\left(2\right)}=\mathcal{S}_P^{\left(2\right)}\,,\nonumber
\ee
with
%
\be
\mathcal{S}_P^{\left(2\right)}=2\frac{M^{2k}}{r^{2k}}\mathcal{E}_2\frac{c_1-c_2 r+\left(3+k\left(2k-3\right)\right)r^2}{\left(r-2M\right)^2}\,,\label{eq:PolarFluidPert}
\ee
with $c_1=2\left(3+4k\left(k-2\right)\right)M^2$ and $c_2=2\left(4+k\left(4k-7\right)\right)M$.

Even though this differential equation admits a solution in closed form, in this case it is simpler to work in terms of Green's functions. The two linearly independent solutions to the homogeneous equation are
\beq
	\Psi_-&=&\frac{3A_1}{M^2}r^2\left(1-\frac{2M}{r}\right) \, , \\
	\Psi_+&=&\frac{A_2}{M^2r\left(r-2M\right)}\bigg(\left(r-M\right)\left(3r^2-6Mr-2M^2\right)M \nonumber\\
	&+&3r^2\left(r-2M\right)^2\text{arctanh}\left(1-\frac{M}{r}\right)\bigg) \, ,
\eeq
and the Wronskian is:
\be
W\left(r\right)\equiv\Psi_+'\left(r\right)\Psi_-\left(r\right)-\Psi_+\left(r\right)\Psi_-'\left(r\right)=\frac{24MA_1A_2}{r\left(2M-r\right)}\, .\nonumber
\ee
Notice that $\Psi_-\left(r\right)$ is regular at the horizon and $\Psi_+\left(r\right)$ at infinity. Imposing the correct boundary conditions enable us to write the solution to the inhomogeneous problem directly,
\beq
H_0^{\left(2\right)}\left(r\right)&=&\Psi_+\left(r\right)\int_{2M}^{r}dr'\, \frac{\mathcal{S}_P^{\left(2\right)}\left(r'\right)\Psi_-\left(r'\right)}{W\left(r'\right)} \nonumber \\
&+&\Psi_-\left(r\right)\int_{r}^{\infty}dr'\, \frac{\mathcal{S}_P^{\left(2\right)}\left(r'\right)\Psi_+\left(r'\right)}{W\left(r'\right)} \,.\label{eq:GreenH0}
\eeq

For $k > 2$ the first integral converges as $r\rightarrow \infty$, and we find that the second one does not contribute to the induced mass quadrupole moment. 
One finds that the TLNs vanish for $k=1$, as expected, but that in general,
\beq
k_2^{E}=\frac{1}{5}\frac{2^{5-2k}\left(2k-1\right)}{2k^2-9k+10}\frac{Q^{2k}}{M^{2k}}=k_2^B \,,\quad k>2\,.
\eeq
Remarkably, the polar TLNs are the same as the axial TLNs, Eq.~\eqref{eq:AxialTLNs_fluid}. This feature was already present in the TLNs of ECOs in the BH limit~\cite{Cardoso:2017cfl}~\footnote{We are grateful to Lam Hui for highlighting this property.}.
A similar procedure can used to obtain the octupolar $l=3$ or higher TLNs.
%
\subsection{Matter away from the horizon: Thin shells} \label{sec:ThinShell}
While the previous results are interesting, astrophysical BHs are thought to have a matter distribution localized away from the horizon.
It is challenging to construct stationary solutions describing astrophysically realistic BH spacetimes. As a surrogate for those setups, we will simply pack all the interstellar material, accretion disk or dark matter in a (infinitesimally) thin shell surrounding a Schwarzschild BH. The dynamics of thin-shells are a vastly explored subject, both in GR~\cite{Israel:1966rt,Brady:1991np,Poisson:1995sv,Lobo:2005zu,Dias:2010uh,LeMaitre:2019xez} and in modified theories of gravity; we refer the interested reader to Poisson~\cite{Poisson:2009pwt} for a pedagogical introduction to the subject. As physical systems, thin-shells are nothing more than very crude approximations. However, their mathematical description is much simpler than more realistic distributions of matter and they often present the key features of these. Thus, what we lose in accuracy is compensated by what we gain in simplicity, making thin-shells the natural starting point to study three dimensional self-gravitating objects. While there are many studies regarding the stability of thin-shells, little has been made in studying the explicit form of gravitational perturbations in spacetimes containing them~\cite{Pani:2009ss,Leung:1999iq,Leung:1999rh,Barausse:2014tra,Uchikata:2016qku}.

Let us then consider the tidal deformation of a distribution of matter whose metric is again given by the general spherically symmetric line element in Eq.~\eqref{eq:sphericalmetric} with
\beq
\begin{cases}
	F\left(r\right)=\bar{\alpha}\left(1-\frac{2M}{r}\right) \, , \, G\left(r\right)=\frac{\bar{\alpha}}{F\left(r\right)} \, , \,  r<r_0\\
	F\left(r\right)=\left(1-\frac{2M_0}{r}\right) \, \, \, , \, G\left(r\right)=\frac{1}{F\left(r\right)} \, \, , \, r>r_0
\end{cases}	
\eeq
Here, $r_0$ is the radius at which the shell is located, $\bar{\alpha}=\frac{1-2M_0/r_0}{1-2M/r_0}$, $M$ is the BH horizon mass, $M_0$ the ADM mass and for future reference we define the shell energy
\beq
	\delta M &\equiv& M_0-M \, . 
\eeq

\subsubsection{Unperturbed solution}
We start by analyzing the unperturbed configuration. The word line of matter elements of the shell is parametrized by
\beq
	x_{\pm}^\mu=x_{\pm}^\mu\left(y^a\right) \, ,
\eeq
where $y^a$ are the intrinsic coordinate functions of the shell, the subscript $+$ or $-$ refers to, respectively, the coordinate chart used outside and inside the shell, and latin indices denote objects defined on the 3D hypersurface of the shell. We choose the intrinsic coordinates of the shell to be
\beq
	y^{a}= \left(T,\Theta,\Phi \right)\, ,
\eeq
and the unperturbed shell is located at
\beq
	x_{+}^\mu&=& \left(T,r_0,\Theta,\Phi \right)\, , \\
	x_{-}^\mu&=& \left(A\,T,r_0,\Theta,\Phi \right)\, .
\eeq
The constant $A$ reflects a possible time-rescaling so that the proper time of the shell is the same for both the exterior and interior coordinate chart. These two regions have to be matched according to the Darmois-Israel junction conditions, which relates the discontinuities on the metric functions with the matter properties of the thin shell \cite{Israel:1966rt}. The first of these imposes that the induced metric, $\gamma_{ab}$, on the 3D hypersurface defined by the shell is continuous
\beq
	\left[\left[\gamma_{ab}\right]\right]=0\, ,
	\label{eq:DarmoiIsraelFirst}
\eeq
where $\left[\left[...\right]\right]$ denotes a jump on a quantity across the shell
\beq
	\left[\left[E\right]\right] \equiv E\left(r_{0_+}\right) - E\left(r_{0_-}\right) \, .
\eeq

The induced metric can be computed through
\beq
	\gamma_{ab}\equiv g_{\mu\nu}\,e_a^\mu \, e_b^\nu \, ,
\eeq
where $e_a^\mu$ are a set of three linearly independent tangent vectors to the shell given by
\beq
	e^{\mu}_a=\frac{\del x^\mu}{\del y^a} \, .
\eeq

The second junction condition determines the stress-energy tensor the shell, $S_{ab}$, in terms of the jump of the extrinsic curvature $K_{ab}$
\beq
	S_{ab}& = - &\frac{1}{8\pi}\left( \left[\left[K_{ab}\right]\right]-\gamma_{ab}\left[\left[K\right]\right] \right) \, ,
	\label{eq:DarmoiIsraelSecond} \\
	K_{ab}&\equiv& e^\mu_a \, e^\nu_b \, \nabla_\mu n_\nu \, , \\
	K&=&\gamma_{ab}K^{ab} \, ,
\eeq
where $n^\mu$ is the unit normal to the thin shell
\be
n_\mu\,e^\mu_a=0 \,,\qquad  n^\mu n_\mu=1 \, .\label{eq:normalThinShell}
\ee

The first junction equation \eqref{eq:DarmoiIsraelFirst} yields
\beq
	A^2=\frac{F_+\left(r_0\right)}{F_-\left(r_0\right)} \, ,
\eeq
which for our model gives $A=1$. Since the configuration is stationary, we can always rescale time such that this is verified and we assume it hereafter. From the second junction condition \eqref{eq:DarmoiIsraelSecond} we obtain
\beq
	S_{TT}&=&-\frac{1}{4\pi r_0}\left[\left[\frac{F}{\sqrt{G}}\right]\right] \, , 
	\label{eq:StressTensorUnperturbed1}\\
	S_{\Theta\Theta}&=&\frac{r_0}{8\pi}\left[\left[\frac{1}{\sqrt{G}}\right]\right] +\frac{r_0}{16\pi}\left[\left[\frac{F'}{f\sqrt{G}}\right]\right] 	\, ,
	\label{eq:StressTensorUnperturbed2}
\eeq
where the prime denotes a radial derivative.

If we consider the thin-shell to be composed by a perfect fluid, its stress tensor is simply
\beq
	S_{ab}=\left(\sigma+ p \right)u_a u_b + p \, \gamma_{ab} \, ,
\eeq	
where $\sigma$ is the surface energy density, $p$ the surface tension and $u^a$ is the fluid's velocity (normalized as $u^\mu u_\mu = -1$). For the unperturbed configuration, the latter is given by
\beq
	u^a=\left(\frac{1}{\sqrt{F\left(r_0\right)}},0,0\right) \, .
\eeq

Using Eqs.~\eqref{eq:StressTensorUnperturbed1}-\eqref{eq:StressTensorUnperturbed2} the surface energy density and pressure are determined by
\beq
	\sigma = -\frac{1}{4\pi r_0}\left[\left[\frac{1}{\sqrt{G}}\right]\right] \, , \\
	\sigma + 2 p =\frac{1}{8\pi}\left[\left[\frac{F'}{F\sqrt{G}}\right]\right]  \, ,
\eeq
which agrees with previous results on thin-shell dynamics \cite{Uchikata:2016qku,Pani:2009ss,Brady:1991np,Visser:2003ge}.
\subsubsection{Perturbed configuration}
To compute the TLNs of this object we need to derive the junction conditions for the stationary, axisymmetric perturbed configuration when the external tidal field is introduced. First, we perturbe the shell radius by
\beq
	\delta r_\pm =\sum_{l,m} \delta r_\pm^{lm}Y^{lm}\left(\Theta,\Phi \right) \, .
\eeq
The junction condition \eqref{eq:DarmoiIsraelFirst} evaluated at first order in the perturbations yields
\beq
	\left[\left[h_0\right]\right]&=&0 \, , \label{eq:junctionh0}\\
	\left[\left[H_0\right]\right]&=&\left[\left[\frac{\delta r \, F'}{F} \right]\right]\, , \label{eq:junctionH0}\\
	\frac{2}{r_0}\left[\left[\delta r \right]\right]&=& -\left[\left[ K \right]\right]\, , \label{eq:junctionK}
\eeq
where frow now on we are omitting the harmonic indexes $l,m$ in the junction conditions to avoid cluttering. 

To apply the second junction condition, we need to consider perturbations to the surface energy density, $\delta \sigma$, and the surface tension, $\delta p$. These are scalars and hence can be simply expanded as
\be
(\delta \sigma,\delta p)=\sum_{l,m} (\delta \sigma^{lm},\delta p^{lm})Y^{lm}\left(\Theta,\Phi \right) \,.
\ee

Finally, we need to perturb the fluid velocity and the unit normal to the shell. The former is determined by imposing the correct normalization and the stationarity condition as done previously in Sec.~\ref{sec:ShortHair}
\beq
	\delta u^a=\sum_{l,m}\frac{1}{2\sqrt{F}}\left(H_0^{lm}-\frac{F'}{F}\delta r^{lm} ,0,0\right)Y^{lm}  \, ,
\eeq
while the latter is computed using \eqref{eq:normalThinShell}
\beq
	\delta n_{\mu_{\pm}}=\sum_{l,m}\sqrt{G}\left(0,\frac{1}{2}H_2^{lm}\,Y^{lm},-\delta r_{\pm}^{lm}\, Y^{lm}_{,\theta} ,0\right)  \, .
\eeq

The second junction condition \eqref{eq:DarmoiIsraelSecond} gives
\beq
	&&\left[\left[\frac{h_{1}}{\sqrt{G}}\right]\right]=0 \, , \\
	&&\frac{1}{2}\left[\left[\frac{h_0'}{\sqrt{G}}\right]\right]-\frac{2}{r_0}\left[\left[\frac{1}{\sqrt{G}}\right]\right]h_0 \nonumber \\
	&-&\frac{1}{2}\left[\left[\frac{F'}{F\sqrt{G}}\right]\right]h_0= 8\pi\sigma\, h_0 \,. \label{eq:junctionh0'}
\eeq

While the first of these agrees with previous results \cite{Uchikata:2016qku,Pani:2009ss}, as far as we aware, the second equation above has not been presented in this form anywhere.

A good sanity check is to see if we have enough equations to solve the problem. To compute the TLNs for this configuration we have to solve the linearized field equations \textit{inside} and \textit{outside} the shell, and then impose the junction conditions derived to fix the constants of integration. As we will shortly see, $h_1$ vanishes for stationary and axisymmetric perturbations, so combining the former equations with \eqref{eq:junctionh0}, we have $2$ junction conditions for $4$ constants of integration coming from the ODEs for $h_0$ ($2$ from inside the shell and $2$ from outside). One of these is fixed by imposing regularity at the BH horizon. Then, we are able to fix 2 more constants and 1 will remain free. Since an overall factor is irrelevant to compute the axial TLNs \eqref{eq:AxialTLNs}, this means we have enough information to completely determine the tidal deformation of the system.

The polar sector couples to matter perturbations and we find more complicated junction conditions
\beq
&&\left[\left[\frac{H_{1}}{\sqrt{G}}\right]\right]=\left[\left[\sqrt{G}\, \delta r \right]\right]=0 \, , \\
&&\frac{2}{r_0^2}\left[\left[\frac{\delta r}{\sqrt{G}}\right]\right] + \frac{2}{r_0}\left[\left[\frac{H_0}{\sqrt{G}}\right]\right]+\frac{1}{r_0}\left[\left[\frac{H_2}{\sqrt{G}}\right]\right] \nonumber \\
&-&\left[\left[\frac{K'}{\sqrt{G}}\right]\right]+\frac{1}{r_0}\left[\left[\frac{\delta r \, G'}{\sqrt{G^3}}\right]\right]-\frac{2}{r_0}\left[\left[\frac{\delta r F'}{F\sqrt{G}}\right]\right] = \nonumber \\
&=&8\pi\, \delta \sigma+ 8\pi\,\sigma\left(\frac{F' \delta r}{F}-H_0\right) \, ,\\
&&\frac{1}{2\,r_0^2}\left[\left[\frac{\delta r}{\sqrt{G}}\right]\right]-\frac{1}{2 r_0}\left[\left[\frac{H_2}{\sqrt{G}}\right]\right] +\frac{2}{r_0}\left[\left[\frac{K}{\sqrt{G}}\right]\right] \nonumber \\
&&-\frac{1}{4}\left[\left[\frac{H_2 F'}{F\sqrt{G}}\right]\right]+\frac{1}{2}\left[\left[\frac{K F'}{F\sqrt{G}}\right]\right]+\frac{1}{2}\left[\left[\frac{K'}{\sqrt{G}}\right]\right] \nonumber \\
&&-\frac{1}{2}\left[\left[\frac{H_0'}{\sqrt{G}}\right]\right] - \frac{1}{2r_0}\left[\left[\frac{\delta r\, G'}{\sqrt{G^3}}\right]\right] +\frac{1}{r_0}\left[\left[\frac{\delta r F'}{F\sqrt{G}}\right]\right] \nonumber \\
&&+\frac{1}{2}\left[\left[\frac{\delta r F'}{f\sqrt{G}}\right]\right]' = 8\pi\, \delta p + 8\pi\, p\left(K+2\frac{\delta r}{r_0} \right) \, .
\eeq
These have to be complemented with an equation of state that relates $\delta p$ with $\delta \sigma$
\be
\delta p=v_s^2 \, \delta \sigma \, , \qquad v_s^2\equiv\left(\frac{dp}{d\sigma}\right)\Bigr\vert_{\sigma_0} \, .
\ee
For standard matter, $v_s$ is the sound of speed of the fluid and can range between $0<v_s^2<1$. Again, the first two of the above conditions agree with previous results \cite{Uchikata:2016qku} while we could not find the last two written in this manner anywhere.

Let us perform the same sanity check as before. $H_1$ will be identically zero from the field equations. Combining these 4 expressions with \eqref{eq:junctionH0} and \eqref{eq:junctionK}, we have 6 junction conditions for 8 constants (4 from $H_0$, 2 from $\delta r$, 2 from $\delta p$ and 2 from $\delta \sigma$; $K$ is related with $H_0$ by the field equations). Once more, one constant is fixed by imposing regularity of $H_0$ at the BH horizon. Hence, we can determine 6 more constants and are left with one free constant that is irrelevant to compute the polar TLNs \eqref{eq:PolarTLNs}. 
\subsubsection{Axial TLNs}
The exterior spacetime has the form of a Schwarzschild metric so using standard results \cite{Cardoso:2017cfl,Binnington:2009bb,Pani:2009ss}
\beq
h_1^{\text{ext}}&=&0 \, , \\
h_0^{\text{ext}}&=&A_1 r^2 _2F^1\left(1-l,l+2;4;\frac{r}{2M_0}\right) \nonumber\\
&+& A_2 G_{2,0}^{2,2} \left( \frac{r}{2M_0}  \bigg | \begin{matrix}1-l & l+2 \\
                                        		-1 & 2 
                          						\end{matrix} \right) \, ,
\eeq
where $G_{2,0}^{2,2}$ is the Meijer function. The first term of $h_0^{\text{ext}}$ corresponds to the external tidal field and the second to the object's response.

For the interior region, the final equation for $h_0^{\text{int}}$ is similar to that in the exterior, with $M_0$ replaced by $M$
\beq
h_0^{\text{int}''}=\frac{4M-l\left(l+1\right)r}{r^2\left(2M-r\right)}h_0^{\text{int}}
\eeq
Consequently, the solution is of the form above substituting $M$ by $M_0$. Imposing regularity of $h_0$ at $r=2M$ (which is inside the shell) means the term with Meijer function has to vanish
\beq
h_0^{\text{int}}&=&A_3 r^2 _2F^1\left(1-l,l+2;4;\frac{r}{2M}\right) \, ,
\eeq
and $h_1^{\text{int}}=0$.

Now, we can impose the junction conditions derived previously. For $l=2$, the general large-distance behavior of $h_0$ is given by a complicated expression. However, we can analyze it by looking at some interesting regimes.

In the limit where the shell is far away,
\beq
k_2^B=\frac{\delta M}{5M_0}\frac{r_0^4}{M_0^4} \,,\quad r_0\rightarrow \infty \,.
\eeq
Notice that, when the shell disappears, $\delta M\rightarrow 0$, $k_2^B\to 0$. This agrees with the well-known vanishing of the TLNs of a BH~\cite{Binnington:2009bb,Cardoso:2017cfl}. The TLN is proportional to the mass in the shell, as we had found for the ``short-hair'' solution. However, the presence of a length scale $r_0$ now implies that the TLNs are very sensitive to the location of the matter. In fact, the $r_0^4/M^4$ dependence is expected on general dimensional grounds, and from comparison with the TLNs of extended configurations such as boson stars.

In the BH limit, when $M_0\rightarrow M$ and $r\rightarrow 2M$
\be
k_2^B \to \frac{8}{5}\frac{\delta M}{M}\left(\frac{r_0}{M}-2\right) \, ,\label{eq:ShellAxialBHLimit}
\ee
which is also compatible with the result for an isolated BH.

It is also interesting to see the behavior of the system when we start without a BH, i.e. $M= 0$. In this case, one finds the exact result 
\beq
k_2^B&=&\frac{8\,\xi}{10\,\mathcal{C}\left(3-3\mathcal{C}-2\mathcal{C}^2+2\mathcal{C}^3\sqrt{\frac{1}{\xi}}\right)+15\xi\,\log\xi}\,,\nonumber \\
\xi&\equiv& 1-\frac{2M_0}{r_0} \,,\qquad	\mathcal{C}\equiv \frac{M_0}{r_0}\,.\nonumber
\eeq
This result seems to be at odds with the claims of Ref.~\cite{Cardoso:2019rvt} that the general scaling of the TLNs of an ECO in the black-hole limit is $k\sim 1/\log\xi$ (see discussion around Eq. (95)). The proof presented there relies on imposing Robin-type boundary conditions, $a\Psi+b\Psi'=c$, on the Zerilli function $\Psi$, at the surface of the compact object, where $a$, $b$ and $c$ depend on the background spacetime. However, the true scaling goes as $k\propto 1/\left(b+\log\xi\right)$, so if in the black-hole limit, $b$ is diverging  faster than the logarithm, their claim does not hold. Notice that the factor $b$ is related with the term yielding information about the derivatives of the perturbations \textit{at} the boundary. For a thin-shell, the perturbations will not be differentiable at such boundary. Therefore, it is not clear how we can rephrase the boundary conditions imposed in Eqs.~\eqref{eq:junctionh0} and \eqref{eq:junctionh0'}, which relate quantities on both sides of the boundary but which are not well defined \textit{at} it, in terms of Robin-type boundary conditions for which the result of Ref.~\cite{Cardoso:2019rvt} applies.

\subsubsection{Polar perturbations}\label{sec:PolarShell}
For the polar sector the behavior of the perturbations inside and outside the shell is similar to the axial case. They are~\cite{Cardoso:2017cfl,Binnington:2009bb,Pani:2009ss}
\beq
	H_0^{\text{ext}}&=&A_1 P^2_l\left(r/M_0-1\right)+A_2Q^2_l\left(r/M-1\right) \, , \\
	H_0^{\text{int}}&=&A_3 P^2_l\left(r/M-1\right) \, , \\
	H_1^{\text{int}}&=&H_1^{\text{ext}}=0 \, ,
\eeq
where regularity of $H_0^{\text{int}}$ at the BH horizon fixes one of the constants. $K$ is determined by the field equations
\beq
K&=&\frac{\left(4M_i^2+2\left(l^2+l-4\right)M_i r-\left(l^2+l-2\right)r^2\right)H_i}{\left(l^2+l-2\right)\left(2M_i-r\right)r}  \nonumber\\
&+&\frac{2M_i\left(2M_i-r\right)H'_i}{\left(l^2+l-2\right)\left(2M_i-r\right)r}
\eeq
where $i$ labels the interior or exterior solution which correspond, respectively, to $M$ or $M_0$.

We can impose the junction conditions and obtain the polar TLNs. For $l=2$, the large distance behavior of $H_0$ is again given by a complicated expression. However, in the large shell radius pressureless ($v_s=0$) limit, the polar TLN is simply
\beq
k_2^E=\left(1+\frac{M_0}{M}\right)\frac{\,\delta M}{2\,M}\frac{r_0^5}{M_0^5}  \,, \quad r_0 \to\infty\label{eq:ShellPolarBigR}
\eeq
%
%
%
in such a way that $k_2^E$ vanishes when $\delta M\rightarrow 0$, as it should.
We note that there is an important dependence on the speed of sound $v_s$. In consequence, $k_2^E$ is positive for small $v_s$ (in the Newtonian limit), but can become negative at large values of $v_s$~\footnote{Negative TLNs have been found in other models involving infinitesimal thin-shells~\cite{Uchikata:2015yma, Uchikata:2016qku} and extended configurations of ECOs~\cite{Cardoso:2017cfl}. They are usually interpreted as leading to a prolation of the deformed compact object, instead of a more intuitive oblate shape. The actual physical reason behind this is still poorly understood.}.

In the BH limit, $M_0\to M$ and $r_0\to 2M$, we find
\beq
k_2^E \to \frac{8\left(3-8v_s^2\right)}{5}\frac{\delta M}{M}\left(\frac{r_0}{M}-2\right) \, ,\label{eq:ShellPolarBHLimit}
\eeq
which has a similar dependence as the axial case \eqref{eq:ShellAxialBHLimit}. Although the numerical coefficients do not exactly match, as occurred for the short hair and ECOs \cite{Cardoso:2017cfl}, we can attribute this difference to the lack of specification of the equation of state. Actually, it is interesting to note that there is perfect agreement between the $l=2$ axial TLN \eqref{eq:ShellAxialBHLimit} and the corresponding polar one \eqref{eq:ShellPolarBHLimit} when $v_s^2=0.25$, which is in the allowed range for $v_s$. Another remark is that $v_s^2 > 3/8$, the $l=2$ polar TLN becomes negative, in agreement with what we observed for the large $r_0$ limit.

If we start without a BH, i.e. $M= 0$, and analyze now the BH limit $r_0\rightarrow 2M_0$ we obtain 
\beq
k_2^E \to \frac{8}{5\left(9+\sqrt{\frac{2}{\xi}}+4v_s^2+3\log\xi\right)}\,.
\eeq
%
\section{Implications for fundamental physics} \label{sec:EoM}

\subsection{Equations of motion} \label{sec:EoM}
In Sec.~\ref{sec:ThinShell}, we showed that the tidal deformability of a thin shell of matter surrounding a BH scales with 
the fifth power of the shell radius $r_0$. This is an unsurprising result, given that a similar scaling holds for spherical bodies in Newtonian gravity or in general relativity~\cite{poisson_will_2014,Mendes:2016vdr,Cardoso:2017cfl} (see also Appendix~\ref{sec:Newtonian} below). We will take it to be valid for more generic matter distributions. As a consequence, the TLNs diverge when matter is placed sufficiently far away, at $r_0\rightarrow \infty$. Since any BH is surrounded by interstellar medium, other galaxies etc, it might at face value seem like the impact on a GW signal is enormous. This is of course physically unreasonable.
The solution to this conundrum is of course tied to the impact that TLNs have on observables. Consider again the scenario discussed in the Introduction~\ref{sec:Intro}, that of a binary system immersed in matter. There will be regions where the matter density is larger than others but, virtually, one can consider that it extends up to spatial infinity. 
The question to be answered is: what portion of this matter is relevant for the tidal effects in the dynamics of and radiation emitted by the binary? In other words, up to which $r_0$ do we need to consider?

Let us look at the equations of motion of the binary system (two objects of masses $M_1$ and $M_2$ and total mass $M_{\rm tot}=M_1+M_2$) at a Newtonian level. To simplify, consider that both bodies only develop a non-negligible mass quadrupole moment through tidal interactions. Then, the equation of motion for the relative position between the objects, $\bm{r}\equiv\bm{r}_1-\bm{r}_2$, to linear order in the quadrupole moments (i.e. neglecting quadrupole-quadrupole interaction) is \cite{poisson_will_2014, Vines:2010ca}
\beq
\frac{d^2r^j}{dt^2}&=&-\frac{M_{\rm tot}}{r^2}\left(1+\frac{9}{r^5}\left(\lambda_{1}\frac{M_2}{M_1}+\lambda_{2}\frac{M_1}{M_2}\right)\right)n^j \, , \label{eq:EoMBinary}
\eeq
where
\beq
\lambda_i&\equiv& \frac{2}{3}k_{2_i} M_i^5 \, \quad r=|\bm{r}| \, ,\quad	\bm{n}=\frac{\bm{r}}{r} \,,
\eeq
being $k_{2_i}$ the $l=2$ polar TLN of each object.

Simplifying even further, take only object ``1'' to be immersed in matter, the other being ``isolated.'' This immediately fixes $k_{2_2}=0$  \cite{Binnington:2009bb,Cardoso:2017cfl}. Then, inserting our results for the $r_0 \rightarrow \infty$ limit of the $l=2$ polar TLNs of a BH surrounded by a thin-shell \eqref{eq:ShellPolarBigR} in the EoM \eqref{eq:EoMBinary}, we expect a dependence as
\beq
	\frac{d^2r^j}{dt^2} \sim \frac{\delta M}{M_1}\frac{r_0^5}{r^5}\frac{M_2}{M_1}n^j \, ,
	\label{eq:ShellEffect}
\eeq
where we have used that in realistic astrophysical scenarios $\delta M \ll M_0$.

Our discussion of tidal interactions relies on treating them as a perturbation, the external field caused by a body in a region far away from the deformed one. This latter condition fixes immediately $r_0/r \ll 1$. However, this condition might not be sufficient to guarantee the first one. From equation \eqref{eq:EoMBinary} and our results for the asymptotic behavior of  $k_2^E$ in the limit $r_0\rightarrow \infty$ \eqref{eq:ShellPolarBigR}, to treat the tidal terms as perturbations, we can only consider matter in a region around the compact objects such that
\beq 
\frac{r_0}{r} \ll \text{min}\left(1\, , \,\left(\frac{M_1}{\delta M}\frac{M_1}{M_2}\right)^{1/5}\right) \,.
\eeq
Although this does not fix $r_0$ to an unambiguous value, it justifies why the scaling of the TLNs with powers of $r_0$ is not problematic.

%
%
%
%

\subsection{Binaries in astrophysical settings} \label{sec:Astro}
Let us now consider a realistic astrophysical system in which the environment might have a measurable impact. As we saw in
Sec.~\ref{sec:EoM},  the leading order effect of tidal interactions in the dynamics of a binary comes from the polar $l=2$ TLN \eqref{eq:EoMBinary}. We also concluded that to use our results for the TLNs of a thin shell \ref{sec:ThinShell}, we had to consider a length scale $r_0$ for the environment smaller than the typical separation $r$ between the binary objects. Accurate modeling of astrophysical systems is an extremely difficult subject \textit{per se}, and is beyond the scope of this paper to make a detailed discussion of the various BH and accretion disk systems that can occur in nature~\cite{2002apa..book.....F,Barausse:2014tra}. In addition, accretion disks are intrinsically dissipative systems, which leads to a transfer of angular momentum between the disk and the companion, through a phenomena known as tidal heating \cite{1979ApJ...233..857G,1993prpl.conf..749L,Hirata:2010vn,poisson_will_2014}. If this transfer is significant, the adiabatic limit taken to study static tidal deformability breaks down. In fact, tidal heating can be more important for the dynamics of the binary than tidal deformability effects~\cite{Maselli:2017cmm,Cardoso:2019rvt}. Tidal heating eventually becomes the dominant effect in the binary evolution, but this regime would also leave distinguishable orbital imprints in both the disk and companion dynamics. We are taking a conservative viewpoint and isolating and discarding all the effects not related to tidally induced multipoles.

The upcoming LISA detector has a frequency band of $f\in \left[10^{-5},1\right]\, \text{Hz}$. The simplest system we can think of is a binary in circular orbit, for which the binary separation $r$ is related with the GW frequency by 
\be
r \sim \left(\frac{GM_{\rm tot}}{\left(\pi f\right)^2}\right)^{1/3}\, .
\ee
Consequently, the lower bound of the LISA frequency band corresponds to binaries separated by $r \sim 10^{6} \left(M_{\rm tot}/M_{\odot} \right)^{1/3}$ km. 

To obtain the properties of the environmental matter, we can use the steady-state model of a Shakura-Sunyaev thin accretion disk~\cite{Shakura:1972te, Abramowicz:2011xu, 2002apa..book.....F,Barausse:2014tra}. This is an axisymmetric, vertically thin disk, i.e. $H<r$ being $H$ the height of the disk. The properties of the disk depend on which matter dominates the pressure (e.g. \textit{gas} or \textit{radiation}) and how opacity is described (\textit{Kramer's law} or \textit{electron scattering}). 

Following Ref.~\cite{Barausse:2014tra}, we parametrize the mass accretion rate with the mass Eddington ratio $f_{\text{Edd}}$, which for thin disks varies between $\sim 10^{-2} \leq f_{\text{Edd}} \leq \sim 0.2 $. This enable us to write the surface density of the thin disk $\Sigma_ {\text{disk}}$ and the disk height $H$ as 
\begin{widetext}
\beq
	\Sigma_ {\text{disk}}\left(r\right) &\approx& 7 \times 10^8 \frac{f_{\text{Edd}}^{7/10}}{\tilde{r}^{3/4}}\left(1-\sqrt{\frac{\tilde{r}_{\text{in}}}{\tilde{r}}} \right)^{7/10}\left(\frac{0.1}{\alpha} \right)^{4/5}\left(\frac{M}{10^6 M_{\odot}} \right)^{1/5} \text{kg}\cdot \text{m}^{-2} \, , \\
	\frac{H}{GM/c^2}\left(r\right) &\approx &3\times10^{-3} f_{\text{Edd}}^{3/20}\left(1-\sqrt{\frac{\tilde{r}_\text{in}}{\tilde{r}}} \right)^{3/20}\left(\frac{0.1}{\alpha}\right)^{1/10}\left(\frac{10^6 M_{\odot}}{M}\right)^{1/10} \tilde{r}^{9/8} \, ,
\eeq
\end{widetext}
where $M$ is the mass of the accreting object, $\tilde{r}=r/\left(GM/c^2\right)$, $\alpha \sim 0.01\, -\, 0.1$ is the viscosity parameter and $\tilde{r}_{\rm in}\sim 6$ is the radius of the inner edge of the disk. The total mass of the disk is then given by
\beq
\delta M &\approx& 2\pi \int_{r_{\rm in}}^{r_{\rm out}}  \Sigma_ {\text{disk}}  r\,dr \,  ,
\eeq
where $r_{\rm out}$ is the radius of the outer edge of the disk. 

To make use of our expression for the $l=2$ polar TLN of the thin shell we have to consider that all of the disk mass is concentrated on a 2-sphere at a radius $r_0$. We consider this $r_0$ to be given by the following average
\be
r_0 =\frac{2\pi}{M} \int_{r_{\text{in}}}^{r_{\text{max}}}  \Sigma_ {\text{disk}}  r^2\,dr \, .
\ee
Finally, the speed of sound $v_s$ is related to the height of the disk via
\be
H \sim \frac{v_s r}{v_K} \, , 
\ee
where $v_K\approx \left(GM/r\right)^{1/2}$ is the local Keplerian velocity.

\subsection{On the minimum measurable TLN}
%
\begin{figure}[t]
\includegraphics[width=0.49\textwidth]{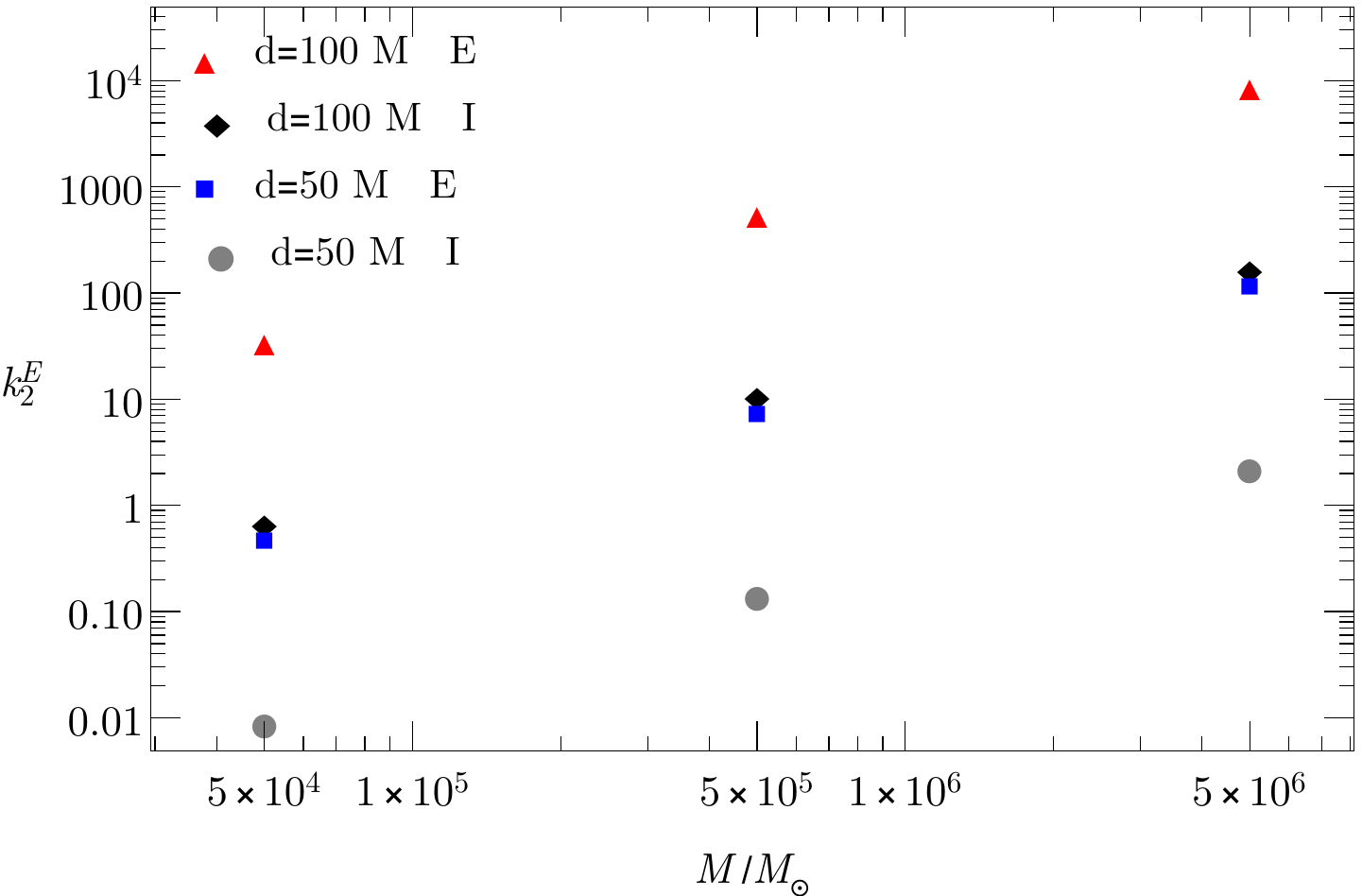}
\caption{$k_2^E$ for a central BH of mass $M$ surrounded by a Shakura-Sunayev thin accretion disk in a circular binary and a companion of the same mass. We present two different binary distances $d$ for $M/M_\odot=5\times 10^4$, $M/M_\odot=5\times 10^5$ and  $M/M_\odot=5\times 10^6$. For each $d$ we present estimations in the most efficient accretion scenario (labeled by $E$), $f_{\text{Edd}}=0.2$ and $\alpha=0.01$, and in the most inefficient one (labeled by $I$)  $f_{\text{Edd}}=0.01$ and $\alpha=0.1$.}
\label{fig:LoveAccretion}
\end{figure}
Now, using the results from \eqref{sec:PolarShell} we can compute the dominant TLN $k_2^E$ of a ``dirty'' BH. Figure~\ref{fig:LoveAccretion} shows $k_2^E$ for representative values of the distance $d=(50\,, 100)M$ and for two different accretion scenarios, an efficient (``E'') with $f_{\text{Edd}}=0.2,\,\alpha=0.01$, and an inefficient (``I'') one with $f_{\text{Edd}}=0.01,\,\alpha=0.1$. This is one of the main results of this work.

The lesson to be learned from Fig.~\ref{fig:LoveAccretion} is that massive objects are typically surrounded by enough
matter that they are perceived as having TLNs of order $\gtrsim 1$. Thus, extreme care and account of environmental
effects has to be taken into account when inferring the properties and nature of ultracompact objects, from a measurement of TLNs~\cite{Cardoso:2017cfl,Maselli:2017cmm,Maselli:2018fay}. This is specially important for extreme-mass-ratio systems, where the long time in band would ideally allow for extremely precise constraints on the TLN of BHs~\cite{Pani:2019cyc}.

We focused on the calculation of an effective TLN, but there are other, possibly dominant effects
in astrophysical environments, such as accretion, gravitational drag~\cite{Barausse:2014tra,Cardoso:2019rou}, or tidal heating effects~\cite{Maselli:2017cmm,Cardoso:2019rvt,Datta:2019epe}.

\section*{Acknowledgements}
%
We thank Sam Gralla for useful feedback and an anonymous referee for valuable suggestions.
V. C. is partially funded by the Van der Waals Professorial Chair.
V.C. acknowledges financial support provided under the European Union's H2020 ERC 
Consolidator Grant ``Matter and strong-field gravity: New frontiers in Einstein's 
theory'' grant agreement no. MaGRaTh--646597. 
F.D. acknowledges financial support provided by FCT/Portugal through grant No. SFRH/BD/143657/2019.
The authors would like to acknowledge networking support by the GWverse COST Action 
CA16104, ``Black holes, gravitational waves and fundamental physics.''
%

\appendix

\section{Newtonian Shell \label{sec:Newtonian}} 

Here, we study the tidal deformability of a spherical shell of matter~\cite{Vogt:2010ad} in Newtonian gravity. This exercise is mainly a sanity check of the results of Sec.~\ref{sec:ThinShell} for the relativistic shell, particularly for the polar TLN when the shell is placed far away from the BH~\eqref{eq:ShellPolarBigR}.

Let us start with a short review of the general theory of tidally deformed compact bodies in Newtonian gravity, based on the pedagogical treatment by Poisson and Will~\cite{poisson_will_2014} and other shorter reviews~\cite{LandryThesis,GuilhermeThesis}. The object is assumed to be formed by a perfect fluid with matter density $\rho$, pressure $p$ and velocity $\textbf{u}$. Consequently, it obeys the Poisson-Euler system
\beq
\nabla^2\Phi &=&- 4\pi G\rho	\, ,\label{eq:PoissonEq} \\
\rho \frac{d\textbf{u}}{dt} &=& \rho \nabla \Phi-\nabla p \, ,\label{eq:Euler}
\eeq
where $G=6.67\times 10^{-11} Nm^2/kg^2$ is Newton's gravitational constant. This system is further complemented with a mass continuity equation
\beq
\frac{\del\rho}{\del t}+\nabla\cdot\left(\rho\,\textbf{u}\right)=0 \, .	\label{eq:Masscontinuity}
\eeq
The isotropy inherent to a perfect fluid implies that, in equilibrium, it is spherically symmetric and we can use spherical coordinates, $\left(r,\theta,\varphi\right)$, centered at the body's center of mass.

Now, introduce an external tidal field, $V$, that perturbs the body's equilibrium configuration. As before, we assume the regime of static tides. Hence, time derivatives are trivial in both Eq.~\eqref{eq:Euler} and \eqref{eq:Masscontinuity}, and the condition of hydrostatic equilibrium is simply 
\beq
\nabla p=\rho \nabla \Phi \, .	\label{eq:HydroEq}
\eeq

In order to exploit the spherical symmetry of the system, it is useful to define the mass function
\be
\frac{dm}{dr}\equiv 4\pi r^2 \rho \, ,
\ee
and rearrange Eq.~\eqref{eq:HydroEq} as 
\be
\frac{dp}{dr}=-\rho \frac{Gm}{r^2} \, .
\ee

For the condition of hydrostatic equilibrium to hold, the tidal field has to be sufficiently far away from the central body. In fact, we will assume that it is placed in vacuum and, therefore, it obeys Laplace's equation
\be
\nabla^2 V=0 \,.	\label{eq:Laplace}
\ee

As in the GR framework, the external field induces deformations in the body's internal structure which are encapsuled in its multipole moments. The tidal Love numbers are again defined as the constant of proportionality between the tidal field multipole moments and the deformed body ones. They only depend on the properties of the deformed body, i.e. its equation of state.

In the most general scenario, one would have to use symmetric tracefree tensors expansions. However, the backgrounds we are interested in are spherically symmetric and we directly employ spherical harmonics~\eqref{eq:SphericalHarmonic} expansions (these two descriptions are equivalent and we redirect the interested reader to Refs.~\cite{LandryThesis,GuilhermeThesis}, where the one-to-correspondence between them is proved).

Returning to the problem, Laplace's equation \eqref{eq:Laplace} is solved by
\be
V=\sum_{lm}\frac{4\pi}{2l+1}d_{lm}r^lY^{lm}\left(\theta,\phi\right)\, ,
\ee
where $d_{lm}$ are called the \textit{tidal moments}. At this point, fluid perturbations are introduced. We follow a surface of constant density, $\rho_0$, which in the unperturbed configuration is at radius $r_0$. Then, we need to consider perturbations in the mass density, $\delta \rho$, and in the radius of such surface, $\delta r$. Once again, Ref.~\cite{poisson_will_2014} offers a pedestrian handling of this, distinguishing between Euler/macroscopic perturbations and Lagrangian/microscopic ones. The former compare quantities at the same position in the space while the latter relates changes on the same fluid element, i.e. we follow a fluid element as it is perturbed. Here, we will skips details of how to handle the differences between these descriptions and directly state that if we follow a spherical surface of matter of density $\rho$ in the microscopic description, the following macroscopic statements are true
\beq
\delta \rho\left(r,\theta,\varphi\right) &=& -\rho' \delta r \left(r,\theta,\varphi\right)\, , \label{eq:DensityFrac}\\
\delta p\left(r,\theta,\varphi\right) &=& -p' \delta r\left(r,\theta,\varphi\right)   \, , \label{eq:PressureFrac}
\eeq
where $\delta r\left(r,\theta,\varphi\right)$ is the change in the radius of the spherical surface of matter. 

The fluid perturbations will change the body's gravitational potential, $\delta \Phi$, so that a perturbed Poisson equation holds
\be
\nabla^2 \delta \Phi = -4\pi G \delta \rho \, .\label{eq:PerturbedPoisson}
\ee

Outside the body, where $\delta \rho $ is zero, this equation is solved by
\be
\Phi^{out}_{lm}=\frac{4\pi G}{2l+1}\frac{I_{lm}}{r^{l+1}} \, ,\label{eq:ExternalPotentialNewton}
\ee
where $I_{lm}$ are the body's multipole moments. Finally, we define the TLNs as
\be
k_l \equiv \frac{1}{2}\left(\frac{c^2}{G\,M}\right)^{2l+1}\frac{G\,I_{lm}}{d_{lm}} \, ,
\ee
where $M$ is the total mass of the object
\be
M\equiv \lim_{r\rightarrow \infty} m\left(r\right) \, ,	\label{eq:Mass}
\ee
and $c$ is the speed of light. Although the multipole moments defined in this appendix are different from the ones used in the main text this definition of TLNs is consistent with the relativistic one for the Polar sector \eqref{eq:PolarTLNs} (there are no Newtonian analogous axial TLNs). Notice also that our definition of $k_l$ differs from that used in Ref.~\cite{poisson_will_2014} as
\be
k_{l_{\text{ours}}} =\left(\frac{c^2\, R}{G\,M}\right)^{2l+1}k_{l_\text{Poisson}} \, .\label{eq:LoveNewton}
\ee
The reasons were explained in relation to Eq.~\eqref{eq:TLNdiff}. 

In order to compute the TLNs \eqref{eq:LoveNewton}, Eq.~\eqref{eq:PerturbedPoisson} has to first be solved inside the body and then the internal and external potential perturbations have to be matched at the body's surfaces. To solve the internal problem, we start by decomposing every perturbation in spherical harmonics
\beq
\delta r &=& \sum_{lm} r f_{lm}\left(r\right)Y_{lm}\left(\theta,\phi\right) \, , \label{eq:RadiusExpansion} \\
\delta X&=& \sum_{lm}\delta X_{lm}\left(r\right)Y_{lm}\left(\theta,\phi\right) \,, \label{eq:PotentialExpansion}
\eeq
with $X=\rho, p, \Phi$ or $V$. Inserting these in Eq.~\eqref{eq:PerturbedPoisson} yields
\be
r^2 \delta \Phi_{lm}''+2r \delta \Phi_{lm}'-l\left(l+1\right)\delta \Phi_{lm}=-4\pi G r^2 \delta \rho_{lm} \,,	\label{eq:PoissonPert}
\ee
while Euler's Eq.~\eqref{eq:Euler} expanded to first order gives
\be
\frac{\delta \rho}{\rho^2}\del_ip-\frac{1}{\rho}\del_i\delta p+\del_i\left(\delta \Phi+V\right)=0 \, ,	\label{eq:EulerPert}
\ee
where $i$ labels the coordinate. Making use of the spherical hamonic decompositions \eqref{eq:RadiusExpansion}-\eqref{eq:PotentialExpansion}, Eq.~\eqref{eq:EulerPert} gives for the radial and angular components, respectively,
\beq
\delta p'_{lm}&=&-\frac{G m}{r^2}\delta \rho_{lm}+\rho\left(\delta \Phi'_{lm}+ V'_{lm}\right) \, , \\
\delta p_{lm}&=&\rho\left(\delta \Phi_{lm}+ V_{lm}\right) \, .
\eeq
Differentiating the latter equation and inserting it in the former, and using Eq.~\eqref{eq:DensityFrac} and \eqref{eq:PressureFrac} gives the following equality
\be
\frac{Gm}{r}f_{lm}=\delta \Phi_{lm}+ V_{lm} \, .	\label{eq:RelationModes}
\ee

Finally, we match this expression with the external one \eqref{eq:ExternalPotentialNewton} at the surface of the body. This is accomplished by remembering that the gravitational potential has to be smooth. In practice, one imposes continuity of $\delta \Phi$ and it first derivative. Using the linear relation given by Eq.~\eqref{eq:LoveNewton} the expression for the TLNs is obtained
\be
k_l=\left(\frac{c^2\,R}{G\,M}\right)^{2l+1}\frac{l+1-\eta_l\left(R\right)}{2\left(l+\eta_l\left(R\right)\right)} \, , 	\label{eq:TLNRad}
\ee	
where $\eta_l$ is dubbed Radau's function
\be
\eta_l\left(r\right)\equiv \frac{r f_{lm}'\left(r\right)}{f_{lm}\left(r\right)} \, .
\ee	

These results indicate that the fractional deformation modes $f_{lm}$ completely determine the structure of the tidally deformed body. To compute them, we transform the ODE for $\Phi_{lm}$ \eqref{eq:PoissonPert} into one for $f_{lm}$ by making use of Eq.~\eqref{eq:RelationModes}
\be
r^2f_{lm}''+6\mathcal{D}\left(r\right)\left(rf'_{lm}+f_{lm}\right)-l\left(l+1\right)f_{lm}=0\, ,	\label{eq:Clairaut}
\ee
where
\be
\mathcal{D}\left(r\right)\equiv \frac{4\pi\rho\left(r\right)r^3}{3m\left(r\right)} \, .
\ee

Notice that it is this function $\mathcal{D}\left(r\right)$ that contains the information about the internal structure of the deformed body, namely it depends on its equation of state. For objects that possess an hard surface, this treatment also allows to determine the geometrical shape of the deformed boundary, which is now described by $R+\delta R\left(\theta,\phi\right)$. Since this is a surface of constant density, it obeys the same equations above for $\delta r$. From \eqref{eq:RelationModes} evaluated at the boundary, we conclude that $\delta R$ depends linearly on the external potential. Expanding $\delta R$ in spherical harmonics
\be
\delta R = \sum_{lm}\frac{4\pi}{2l+1} \delta R_{lm}\left(r\right)Y_{lm}\left(\theta,\phi\right) \, , 
\ee
allow us to introduce a new Love number that fully characterizes the shape of the tidally deformed object
\be
h_l = \frac{G\,M}{R^{l+2}}\frac{\delta R_{lm}}{d_{lm}} \, .
\ee

This is called the \textit{surficial Love number} and we normalize it as in Ref.~\cite{poisson_will_2014}, since this last discussion only makes sense for objects with an hard surface. Finally, the matching gives
\be
h_l=\frac{2l+1}{l+\eta_l\left(R\right)} \, .	\label{eq:SurficialTLN}
\ee

Now, we will solve this problem for a model of a spherical shell given by Vogt and Letelier \cite{Vogt:2010ad}. This is represented by the gravitational potential and matter density
\beq 
	\Phi\left(r\right) &=& -\frac{G M}{\left(r^n+r_0^n\right)^{1/n}}	\, , \label{eq:ShellPotential}\\
	\rho\left(r\right) &=& \frac{M\left(n+1\right)b^nr^{n-2}}{4\pi\left(r^n+r_0^n\right)^{2+1/n}} \, , \label{eq:ShellDensity}
\eeq 
where $r_0$ is a parameter with units of length, $M$ is the total mass of the shell \eqref{eq:Mass} and $n>0$. For $n>2$, $\rho$ vanishes at $r=0$ and the mass distribution indeed represents a shell. As $n$ increases, the shell becomes thinner and localized around $r=r_0$. In the limit $n\rightarrow \infty$ this model describes an infinitesimal thin shell located at $r=r_0$.

The formalism developed to compute the TLNs relies on making a match at the surface of the compact object. However, this shell does not possess an hard surface. A possible solution to this problem occurs if the matter density is sufficiently localized so that the matching is well defined in the limit $R\rightarrow \infty$. This is the scenario that occurs in boson stars, whose tidal deformations in both Newtonian gravity and general relativity were studied in \cite{Mendes:2016vdr} (Ref.\cite{Cardoso:2017cfl} complements their work in general relativity). However, for them the matter distribution decays exponentially while here it happens slower so there is no guarantee \textit{a priori} that the problem is well posed.

The solution of Eq.\eqref{eq:Clairaut} which is regular at $r=0$ is
\be
f_{lm}\left(y\right)=c_1 \, y^{-d}\, _2\widetilde{F}^1\left(a,b,1+\frac{c}{n};-y^n\right) \, ,	\label{eq:SolutionNewtonShell}
\ee	
where 
\beq
	y&\equiv&\frac{r}{r_0} \, , \\
	a&=&-1-\frac{l}{n}+\frac{c-3}{2n} \, ,\\
	b&=&-1+\frac{l}{n}+\frac{c-1}{2n} \, , \\
	c&=&\sqrt{-7+4n\left(n-1\right)+4l\left(l+1\right)} \, ,\\
	d&=&n+\frac{1-c}{2}  \, ,
\eeq	
and $_2\widetilde{F}^1\left(a,b,c;x\right)$ are the regularized hypergeometric functions \cite{AbraSteg72}. The TLNs of the shell can then be computed by plugging this solution in Eq.~\eqref{eq:TLNRad} and taking the limit $R\rightarrow \infty$. We conclude that the solution only converges for $n>2l+1$. For smaller values of $n$, the equality \eqref{eq:RelationModes} is not respected in the $R \rightarrow \infty$ limit. 

When the problem is well posed we find
\begin{widetext}
\beq
		k_l	= -\frac{\left(1+n\right)\left(1-2n+c\right)}{2\,n^2\left(b+1\right)}\frac{\Gamma\left(a-b\right)}{\Gamma\left(b-a\right)}\frac{\Gamma\left(b\right)}{\Gamma\left(a+1\right)}\frac{\Gamma\left(2+\frac{1-2l}{n}+b\right)}{\Gamma\left(2+\frac{3+2l}{n}+a\right)}\left(\frac{c^2\, r_0}{G\,M}\right)^{2l+1} \, , \, n>2l+1		\label{eq:TLNNewtonShell}
\eeq
\end{widetext}


The upshot is that the TLNs of this Newtonian shell are of order $\mathcal{O}\left(\frac{c^2\, r_0}{G\,M}\right)^{2l+1}$ and bounded below in the thin-shell limit by
\beq
\lim_{n\rightarrow \infty} k_l	= \frac{l+2}{2\left(l-1\right)} \left(\frac{c^2\,r_0}{G\,M}\right)^{2l+1}\, .	\label{eq:ThinShellNewtonian}
\eeq
Notice that for $l=2$ we obtain the scaling $k_2\propto r_0^{5}$, important for the discussion in the main text.
In this limit we can also compute the surficial Love number \eqref{eq:SurficialTLN}
\beq
\lim_{n\rightarrow \infty} h_l	= \frac{2l+1}{l-1} \, .
\eeq

\bibliographystyle{h-physrev4}
\bibliography{references} 

\end{document}